\documentclass[twocolumn]{aa}
\usepackage{graphicx,epsf}
\usepackage[figuresright]{rotating}
\begin{document}
\renewcommand{\topfraction}{0.85}
\renewcommand{\bottomfraction}{0.7}
\renewcommand{\textfraction}{0.15}
\renewcommand{\floatpagefraction}{0.90}
   \title{Observations of Selected AGN with H.E.S.S.}

\author{F. Aharonian\inst{1}
 \and A.G.~Akhperjanian \inst{2}
 \and A.R.~Bazer-Bachi \inst{3}
 \and M.~Beilicke \inst{4}
 \and W.~Benbow \inst{1}
 \and D.~Berge \inst{1}
 \and K.~Bernl\"ohr \inst{1,5}
 \and C.~Boisson \inst{6}
 \and O.~Bolz \inst{1}
 \and V.~Borrel \inst{3}
 \and I.~Braun \inst{1}
 \and F.~Breitling \inst{5}
 \and A.M.~Brown \inst{7}
 \and P.M.~Chadwick \inst{7}
 \and L.-M.~Chounet \inst{8}
 \and R.~Cornils \inst{4}
 \and L.~Costamante \inst{1,20}
 \and B.~Degrange \inst{8}
 \and H.J.~Dickinson \inst{7}
 \and A.~Djannati-Ata\"i \inst{9}
 \and L.O'C.~Drury \inst{10}
 \and G.~Dubus \inst{8}
 \and D.~Emmanoulopoulos \inst{11}
 \and P.~Espigat \inst{9}
 \and F.~Feinstein \inst{12}
 \and G.~Fontaine \inst{8}
 \and Y.~Fuchs \inst{13}
 \and S.~Funk \inst{1}
 \and Y.A.~Gallant \inst{12}
 \and B.~Giebels \inst{8}
 \and S.~Gillessen \inst{1}
 \and J.F.~Glicenstein \inst{14}
 \and P.~Goret \inst{14}
 \and C.~Hadjichristidis \inst{7}
 \and M.~Hauser \inst{11}
 \and G.~Heinzelmann \inst{4}
 \and G.~Henri \inst{13}
 \and G.~Hermann \inst{1}
 \and J.A.~Hinton \inst{1}
 \and W.~Hofmann \inst{1}
 \and M.~Holleran \inst{15}
 \and D.~Horns \inst{1}
 \and A.~Jacholkowska \inst{12}
 \and O.C.~de~Jager \inst{15}
 \and B.~Kh\'elifi \inst{1}
 \and Nu.~Komin \inst{5}
 \and A.~Konopelko \inst{1,5}
 \and I.J.~Latham \inst{7}
 \and R.~Le Gallou \inst{7}
 \and A.~Lemi\`ere \inst{9}
 \and M.~Lemoine-Goumard \inst{8}
 \and N.~Leroy \inst{8}
 \and T.~Lohse \inst{5}
 \and J.M.~Martin \inst{6}
 \and O.~Martineau-Huynh \inst{16}
 \and A.~Marcowith \inst{3}
 \and C.~Masterson \inst{1,20}
 \and T.J.L.~McComb \inst{7}
 \and M.~de~Naurois \inst{16}
 \and S.J.~Nolan \inst{7}
 \and A.~Noutsos \inst{7}
 \and K.J.~Orford \inst{7}
 \and J.L.~Osborne \inst{7}
 \and M.~Ouchrif \inst{16,20}
 \and M.~Panter \inst{1}
 \and G.~Pelletier \inst{13}
 \and S.~Pita \inst{9}
 \and G.~P\"uhlhofer \inst{1,11}
 \and M.~Punch \inst{9}
 \and B.C.~Raubenheimer \inst{15}
 \and M.~Raue \inst{4}
 \and J.~Raux \inst{16}
 \and S.M.~Rayner \inst{7}
 \and A.~Reimer \inst{17}
 \and O.~Reimer \inst{17}
 \and J.~Ripken \inst{4}
 \and L.~Rob \inst{18}
 \and L.~Rolland \inst{16}
 \and G.~Rowell \inst{1}
 \and V.~Sahakian \inst{2}
 \and L.~Saug\'e \inst{13}
 \and S.~Schlenker \inst{5}
 \and R.~Schlickeiser \inst{17}
 \and C.~Schuster \inst{17}
 \and U.~Schwanke \inst{5}
 \and M.~Siewert \inst{17}
 \and H.~Sol \inst{6}
 \and D.~Spangler \inst{7}
 \and R.~Steenkamp \inst{19}
 \and C.~Stegmann \inst{5}
 \and J.-P.~Tavernet \inst{16}
 \and R.~Terrier \inst{9}
 \and C.G.~Th\'eoret \inst{9}
 \and M.~Tluczykont \inst{8,20}
 \and G.~Vasileiadis \inst{12}
 \and C.~Venter \inst{15}
 \and P.~Vincent \inst{16}
 \and H.J.~V\"olk \inst{1}
 \and S.J.~Wagner \inst{11}}
 
\institute{
Max-Planck-Institut f\"ur Kernphysik, Heidelberg, Germany;
\and
Yerevan Physics Institute, Armenia;
\and
Centre d'Etude Spatiale des Rayonnements, CNRS/UPS, Toulouse, France;
\and
Universit\"at Hamburg, Institut f\"ur Experimentalphysik, Germany;
\and
Institut f\"ur Physik, Humboldt-Universit\"at zu Berlin, Germany;
\and
LUTH, UMR 8102 du CNRS, Observatoire de Paris, Section de Meudon, France;
\and
University of Durham, Department of Physics, U.K.;
\and
Laboratoire Leprince-Ringuet, IN2P3/CNRS,
Ecole Polytechnique, Palaiseau, France;
\and
APC, Paris, France;
\thanks{UMR 7164 (CNRS, Universit\'e Paris VII, CEA, Observatoire de Paris)}
\and
Dublin Institute for Advanced Studies, Ireland;
\and
Landessternwarte, K\"onigstuhl, Heidelberg, Germany;
\and
Laboratoire de Physique Th\'eorique et Astroparticules, IN2P3/CNRS,
Universit\'e Montpellier II, France;
\and
Laboratoire d'Astrophysique de Grenoble, INSU/CNRS, Universit\'e Joseph Fourier, France;
\and
DAPNIA/DSM/CEA, CE Saclay, Gif-sur-Yvette, France;
\and
Unit for Space Physics, North-West University, Potchefstroom, South Africa;
\and
Laboratoire de Physique Nucl\'eaire et de Hautes Energies, IN2P3/CNRS, Universit\'es
Paris VI \& VII, France;
\and
Institut f\"ur Theoretische Physik, Lehrstuhl IV, Ruhr-Universit\"at Bochum, Germany;
\and
Institute of Particle and Nuclear Physics, Charles University, Prague, Czech Republic;
\and
University of Namibia, Windhoek, Namibia;
\and
European Associated Laboratory for Gamma-Ray Astronomy, jointly
supported by CNRS and MPG;}
 
   \date{Received 20 May 2005; Accepted 16 June 2005}

   \abstract{
A sample of selected active galactic nuclei (AGN) was observed in 2003 and 2004
with the High Energy Stereoscopic System (H.E.S.S.), an array of imaging atmospheric-Cherenkov
telescopes in Namibia.   The redshifts of these candidate very-high-energy (VHE, $>$100 GeV) 
$\gamma$-ray emitters range from $z$=0.00183 to $z$=0.333.  Significant detections were already
reported for some of these objects, such as PKS\,2155$-$304 and Markarian\,421. 
Marginal evidence (3.1$\sigma$) for a signal is found 
from large-zenith-angle observations of Markarian\,501, corresponding to an integral flux of
I($>$1.65 TeV) = (1.5$\pm$0.6$_{\mathrm{stat}}$$\pm$0.3$_{\mathrm{syst}}$) 
$\times$ 10$^{-12}$ cm$^{-2}$ s$^{-1}$ or $\sim$15\% 
of the Crab Nebula flux.  Integral flux upper limits for 19 other AGN, based on exposures
of $\sim$1 to $\sim$8 hrs live time, and with average energy thresholds 
between 160 GeV and 610 GeV, range from 0.4\% to 5.1\% of the Crab Nebula flux.  
All the upper limits are the most constraining ever reported for these objects.

   }

   \maketitle
%

\section{Introduction}

Active galactic nuclei are known to emit radiation over the entire electromagnetic spectrum,
from radio waves to TeV $\gamma$-rays.  These objects, which are found in only a small fraction of
the total number of observed galaxies, are very luminous, extremely compact, and
can exhibit large luminosity variations on time scales ranging from less than an hour up to 
several years.  Although AGN differ widely in their observed characteristics, 
a unified description (as reviewed in \cite{AGN_model}) has emerged in which an AGN consists of 
a super-massive black hole ($10^7-10^{10}$ solar masses) surrounded in the inner
regions by an accretion disk, and in the outer regions by a thick torus of gas and dust.  
In some AGN (the radio-loud population, $\sim$10\%), a highly relativistic outflow of energetic particles 
exists approximately perpendicular to the accretion disk and torus plane.  
This flow forms collimated radio-emitting jets which generate 
the non-thermal emission observed from radio to $\gamma$-rays.
It is believed that some of the numerous AGN classifications result from viewing these objects at
various orientation angles with respect to the torus plane.  Essentially all
AGN detected at VHE energies (shown in order of redshift with references in Table~\ref{VHE_AGN}) 
are radio-loud objects of the BL\,Lacertae (BL Lac) type, which have their jet pointed
close to the observer's line of sight.  An exception to this exists with the detection 
of the Fanaroff-Riley type I radio galaxy M\,87 at VHE energies. Although M\,87 is believed
to be a mis-aligned BL\,Lac \cite{M87_misalign}, it is not clear whether the VHE emission
comes from the jet or the central object.

   \begin{table*}
      \caption{Known VHE AGN in order of redshift, along with the
references for the initial discovery by other VHE instruments
and references for H.E.S.S. results.  Known is defined here as having VHE
detections reported by at least two different instruments or by H.E.S.S..}
         \label{VHE_AGN}
        \centering
         \begin{tabular}{c c c c}
            \hline\hline
            \noalign{\smallskip}
	     AGN & $z$ & Reference & H.E.S.S. Reference\\
            \noalign{\smallskip}
            \hline
            \noalign{\smallskip}
             M\,87          & 0.004 & ~\cite{M87_disc} &  ~\cite{HESS_M87}\\
	     Markarian\,421       & 0.030 & ~\cite{mkn421_disc} & ~\cite{HESS_421}\\
	     Markarian\,501       & 0.034 & ~\cite{mkn501_disc} & \\
	     1ES\,2344+514  & 0.044 & ~\cite{2344_disc} & \\
             1ES\,1959+650  & 0.047 & ~\cite{1959_disc} & \\
	     PKS\,2005-489  & 0.071 & & ~\cite{pks2005_paper} \\
	     PKS\,2155-304  & 0.116 & ~\cite{2155_disc} & ~\cite{pks2155_paper}\\
	     1ES\,1426+428  & 0.129 & ~\cite{1426_disc} & \\
          \noalign{\smallskip}
            \hline
       \end{tabular}
   \end{table*}

The known VHE AGN have helped to constrain significantly the models for production of VHE $\gamma$-rays
through spectral and variability studies.  However, there are still many differing models
that describe the present data, making a larger sample of known VHE AGN necessary to make
more definitive conclusions.  Also, VHE photons are absorbed by interactions
on the extragalactic background light (EBL) leading to an energy dependent
horizon for viewing VHE sources.  The energy spectrum of VHE AGN may exhibit characteristics,
such as steepening of the spectrum and a cutoff, as a result of this absorption.  Interpretation
of such features can be used as a probe of the EBL in the 
optical and near-IR regimes (\cite{EBL_effect}; \cite{EBL_effect2}), 
for which direct measurements are dominated
by large systematic uncertainties.  Since such an interpretation is complicated by
discerning which features are a result of EBL absorption and which are intrinsic to the
object, a large data set of VHE AGN at differing redshifts are needed to ascertain
which effects can be attributed to the EBL.

A large sample of AGN located at z$<$0.333 was observed by H.E.S.S. in 2003 and 2004.
Most of these objects are BL\,Lacs, many of which are suggested as good candidates 
for detection as VHE emitters (\cite{luigi_AGN}; ~\cite{perlman_AGN}; \cite{stecker}).
A sample of nearby non-blazar AGN, like M\,87, was also observed with 
the hope of extending the known VHE-bright AGN to other classes.  These include a set of famous
radio-loud galaxies, characterized by resolved radio, optical and X-ray jets (Cen\,A, Pictor\,A,
3C\,120, and the quasar 3C\,273) and a sample of radio-weak objects (the Seyfert galaxies NGC\,1068, 
NGC\,3783 and NGC\,7469).  The detections resulting from the H.E.S.S. AGN observation 
program have been reported elsewhere (see Table~\ref{VHE_AGN} for references).  These include
the confirmation of the VHE emission seen from M\,87 and PKS\,2155$-$304, the
detection of Markarian\,421 using large-zenith-angle observations,
and the discovery of VHE emission from 
PKS\,2005$-$489.  Flux upper limits, the strongest ever produced, 
from the non-detection of the remaining objects are presented here.  

\section{H.E.S.S. Detector}

The H.E.S.S. experiment, a square array (120 m side) 
of four imaging atmospheric-Cherenkov telescopes located in the Khomas Highland of Namibia
(23$^{\circ}$ 16' 18'' S, 16$^{\circ}$ 30' 1'' E, 1800 m above sea level),
uses stereoscopic observations of $\gamma$-ray induced air showers
to search for astrophysical $\gamma$-ray 
emission above $\sim$100 GeV.  Each telescope has a 107 m$^{2}$ tessellated mirror 
dish and a 5$^{\circ}$ field-of-view (f.o.v.) camera 
consisting of 960 individual photomultiplier pixels.
The sensitivity of H.E.S.S. (5$\sigma$ in 25 hours for a 1\% Crab Nebula flux 
source at 20$^{\circ}$ zenith angle)
allows for detection of VHE emission from objects at previously
undetectable flux levels.  More details on H.E.S.S. can be 
found in ~\cite{HESS1}, ~\cite{cent_trig}, ~\cite{HESS2},  and ~\cite{HESS3}.

\section{Observations}
 
The H.E.S.S. observations of AGN in 2004 use the full four-telescope array. 
For some of the data, individual telescopes were excluded from the observations or analysis
due to hardware problems.  Also,
2003 observations of 1ES\,0323+022 were made prior to the completion of the array and thus
use only two or three telescopes.  While the sensitivity of H.E.S.S. is less during observations
with fewer telescopes, it is still unprecedented.  
Table~\ref{results} shows the candidate AGN observed by H.E.S.S. and
gives details of the observations that pass selection criteria which
remove data for which the weather conditions were poor or the hardware was not
functioning properly.  The data were taken in 28 minute runs using {\it Wobble} mode, i.e. 
the source direction is offset, typically by $\pm$0.5$^{\circ}$, relative 
to the center of the f.o.v. of the camera during observations, which
allows for both on-source observations and simultaneous estimation
of the background induced by charged cosmic rays.  As the energy threshold of H.E.S.S. observations increases
with zenith angle, the mean zenith angle of the exposure for each of the AGN along with the corresponding
average energy threshold (after selection cuts) of those observations is also shown in Table~\ref{results}.
It should be noted that the H.E.S.S. Monte Carlo simulations show that the azimuthal angle at which an 
object is observed has a small effect on the energy threshold of observations.  
Sources which culminate in the south (i.e. those with declination less than the latitude of H.E.S.S.) have 
slightly higher energy thresholds (e.g. compare 1ES\,0323+022 with Pictor\,A).

   \begin{sidewaystable*}
      \caption{The candidate AGN ordered by right ascension in groups of BL\,Lacs, other radio-loud galaxies and
radio-weak galaxies. The coordinates (J2000), redshift, 
and type (BL=BL\,Lacs, FSRQ=Flat Spectrum Radio Quasar, Sy=Seyferts (types I \& II), 
FR=Fanaroff-Rileys (types I \& II)) shown are taken
from the SIMBAD Astronomical Database and the NASA/IPAC Extragalactic Database.  Observations not using
the full array are shown by the number of telescopes (N$_{\mathrm{tel}}$) column along
the breakdown of the number of observation 
runs (N$_{\mathrm{runs}}$) and live time (T).  N$_{\mathrm{tel}}$=2$^*$ refers
to a configuration of two telescopes on opposite sides of the square (170 m separation).  The mean 
zenith angle of observation (Z$_{\mathrm{obs}}$), the corresponding post-selection cuts energy threshold 
(E$_{\mathrm{th}}$), the number of on-source and off-source counts, the off-source normalization, the observed
excess and significance (S) are also shown.}
         \label{results}
        \centering
         \begin{tabular}{c c c c c c c c c c c c c c c}
            \hline\hline
            \noalign{\smallskip}
	     Object & $\alpha$ & $\delta$ & $z$ & Type &  N$_{\mathrm{tel}}$ & N$_{\mathrm{runs}}$ & T & Z$_{\mathrm{obs}}$ & E$_{\mathrm{th}}$ & On & Off & Norm & Excess & S \\
             & [hh mm ss] & [dd mm ss] & & & &  & [hrs] & [$^{\circ}$] & [GeV] & & & & & [$\sigma$] \\
            \noalign{\smallskip}
            \hline
            \noalign{\smallskip}
	{\it BL\,Lacs} \\
            \noalign{\smallskip}
	     1ES\,0145+138    & 01 48 29.8 & +14 02 19 & 0.125 & BL & 3,4 & 12 (4,8) & 4.3 (1.5,2.8) & 40 & 310 & 206 & 2287 & 0.08953 & 1.2 & 0.1 \\
	     1ES\,0229+200    & 02 32 48.6 & +20 17 17 & 0.140 & BL & 4 & 2 & 0.8 & 46 & 410 & 24 & 280 & 0.09077 & $-$1.4 & $-$0.3 \\
	     1ES\,0323+022    & 03 26 14.0 & +02 25 15 & 0.147 & BL & 2$^*$,3,4 & 12 (3,3,6) & 4.7 (1.3,1.1,2.3) & 29 & 210 & 279 & 3009 &0.09034 & 7.2 & 0.4 \\
             PKS\,0548$-$322    & 05 50 40.6 & $-$32 16 16 & 0.069 & BL & 4 & 11 & 4.1 & 20 & 190 & 340 & 3321 & 0.09065 & 39.0 & 2.1 \\
             EXO\,0556.4$-$3838 & 05 58 06.2 & $-$38 38 27 & 0.034 & BL & 4 & 3 & 1.2 & 33 & 250 & 101 & 934 & 0.08968 & 17.2 & 1.7 \\
             RGB\,J0812+026   & 08 12 01.9 & +02 37 33 & --- & BL & 3,4 & 4 (2,2) & 0.7 (0.5,0.2) & 30 & 220 & 44 & 519 & 0.08999 & $-$2.7 & $-$0.4 \\
             RGB\,J1117+202   & 11 17 06.2 & +20 14 07 & 0.139 & BL & 3,4 & 9 (3,6) & 3.8 (1.2,2.6) & 52 & 610 & 155 & 1703 & 0.08990 & 1.9 & 0.1 \\
	     1ES\,1440+122    & 14 42 48.3 & +12 00 40 & 0.162 & BL & 4 & 2 & 0.9 & 38 & 290 & 59 & 633 & 0.08955 & 2.3 & 0.3 \\
             Markarian\,501         & 16 53 52.2 & +39 45 37 & 0.034 & BL & 4 & 4 & 1.8 & 64 & 1650 & 95 & 742 & 0.09017 & 28.1 & 3.1 \\
             RBS\,1888        & 22 43 42.0 & $-$12 31 06 & 0.226 & BL & 4 & 6 & 2.6 & 15 & 170 & 190 & 2330 & 0.08958 & $-$18.7 & $-$1.3 \\
             Q\,J22548$-$2725   & 22 54 53.2 & $-$27 25 09 & 0.333 & BL & 4 & 5 & 2.1 & 13 & 170 & 200 & 2294 & 0.08986 & $-$6.1 & $-$0.4 \\
             PKS\,2316$-$423    & 23 19 05.9 & $-$42 06 49 & 0.055 & BL & 4 & 5 & 2.2 & 21 & 190 & 169 & 2032 & 0.08952 & $-$12.9 & $-$0.9 \\
 	     1ES\,2343$-$151    & 23 45 37.8 & $-$14 49 10 & 0.226 & BL & 3,4  & 6 (2,4) & 2.6 (0.9,1.7) & 11 & 160 & 235 & 2652 & 0.08966 & $-$2.8 & $-$0.2 \\
	\\
	{\it Radio-Loud Galaxies} \\
            \noalign{\smallskip}
	     3C\,120          & 04 33 11.1 & +05 21 16 & 0.033 & FR I & 4 & 12 & 5.0 & 32 & 230 & 232 & 3023 & 0.09087 & $-$42.7 & $-$2.5 \\
             Pictor\,A        & 05 19 49.7 & $-$45 46 45 & 0.034 & FR II & 4 & 18 & 7.4 & 27 & 220 & 516 & 5451 & 0.09041 & 23.2 & 1.0 \\
	     3C\,273          & 12 29 06.7 & +02 03 09 & 0.158 & FSRQ & 4 & 9 & 3.9 & 37 & 280 & 245 & 2634 & 0.08978 & 8.5 & 0.5 \\
             Cen\,A           & 13 25 27.6 & $-$43 01 09 & 0.00183 & FR I & 4 & 10 & 4.2 & 21 & 190 & 465 & 7065 & 0.06461 & 8.5 & 0.4 \\
	\\
	{\it Radio-Weak Galaxies} \\
            \noalign{\smallskip}
             NGC\,1068        & 02 42 40.8 & $-$00 00 48 & 0.00379 & Sy II & 4 & 10 & 4.3 & 27 & 210 & 246 & 2580 & 0.09047 & 12.6 & 0.8 \\
             NGC\,3783        & 11 39 01.8 & $-$37 44 19 & 0.00965 & Sy I & 2$^*$,3,4 & 5 (1,2,2) & 1.8 (0.4,0.8,0.8) & 27 & 220 & 106 & 1205 & 0.08947 & $-$1.8 & $-$0.2 \\
             NGC\,7469        & 23 03 15.8 & +08 52 26 & 0.016 & Sy I & 4 & 10 & 4.3 & 33 & 250 & 271 & 3373 & 0.08962 & $-$31.2 & $-$1.8 \\
          \noalign{\smallskip}
            \hline
       \end{tabular}
   \end{sidewaystable*}

\section{Analysis Technique}

The data passing the run selection criteria are calibrated as detailed in \cite{calib_paper},
and the event reconstruction and background rejection are performed as described in \cite{pks2155_paper},
with some minor improvements discussed in \cite{pks2005_paper}. 
The background is estimated using all events passing selection cuts in a 
number of circular off-source regions offset by the same distance, relative
to the center of the f.o.v., in the sky as 
the on-source region (for more details see \cite{HEGRA_BG}).  
The on-source region, the size of which is optimized for the detection of point sources, 
is a circle of radius $\sim$0.11$^{\circ}$ centered on the source, and each off-source region 
has approximately\footnote{The off-source data are first placed into a pixelated two-dimensional map 
and then integrated in an approximate circle for each region. The difference in total area is of order 1\%.} 
the same area as the on-source region.  
The maximum number of non-overlapping off-source regions fitting in the field of
view are used.  An area around the on-source position, completely 
containing the H.E.S.S. point-spread-function, is excluded to eliminate
possible contamination from poorly reconstructed $\gamma$-rays.  For the typical on-source offset
of 0.5$^{\circ}$, eleven off-source regions are possible.  
In the case of observations of Cen\,A, offset by 
0.7$^{\circ}$, sixteen regions are used.  The statistical error on 
the background measurement is reduced by the use of a larger background sample, and there is no
need for a radial acceptance correction, which accounts for the strongest 
acceptance change across the f.o.v., since the off-source regions
are offset by the same radial distance as the on-source region.  The significance of any excess 
is calculated following the method of Equation (17) in \cite{lima} and all upper limits
are determined using the method of \cite{UL_tech}.

\section{Results}

\begin{figure}
   \centering
      \includegraphics[width=8.7cm]{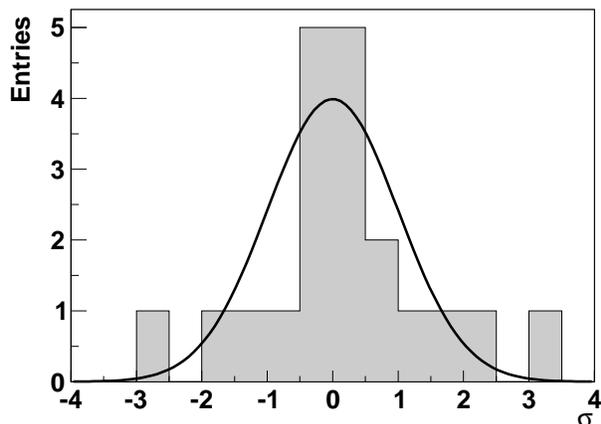} \\ [-0.3cm]
      \caption{Distribution of the significance observed from the 20 selected AGN.  The curve
represents a Gaussian distribution with zero mean and a standard deviation of one.}
         \label{AGN_sigma}
   \end{figure}

Figure~\ref{AGN_sigma} shows the distribution of the significance observed from
the direction of each of the twenty AGN.  No significant excess of VHE $\gamma$-rays is found 
from any of the AGN in the given exposure time ($<$8 hrs each), 
with the possible exception of Markarian\,501 (3.1$\sigma$). Specific details of the results
for each AGN are shown in Table~\ref{results}.  Additionally, a search for serendipitous 
source discoveries in the H.E.S.S. f.o.v. centered on each of the AGN yielded no significant excess.

Given that it is well established that Markarian\,501 is a VHE $\gamma$-ray emitter, the excess (3.1$\sigma$) 
from the only night (MJD 53172) of observations of Markarian\,501 can be treated as significant 
and a flux calculated.  Assuming the spectrum measured above 1.5 TeV by the High Energy
Gamma Ray Astronomy (HEGRA) experiment \cite{HEGRA_501},
a power law with a photon index of $\Gamma$=2.6, the corresponding integral flux above the 1.65 TeV energy
threshold is I($>$1.65 TeV) = (1.5$\pm$0.6$_{\mathrm{stat}}$$\pm$0.3$_{\mathrm{syst}}$) 
$\times$ 10$^{-12}$ cm$^{-2}$ s$^{-1}$ or $\sim$15\% of
the H.E.S.S. Crab Nebula flux above this threshold. While the VHE flux from Markarian\,501 
is known to be highly variable, the measured flux is similar to the value reported in 
Bradbury et al. (1997).

   \begin{table*}
      \caption{Integral flux upper limits (99.9\% confidence level, assuming a power law spectrum with
$\Gamma$=3.0) above the energy
threshold of observations (E$_{\mathrm{th}}$) and the corresponding
percentage of the H.E.S.S. Crab Nebula flux from observations of 
each of the candidate AGN by H.E.S.S.. References to 
upper limits available from other VHE instruments are also shown: HEGRA ~\cite{HEGRA_AGN},
Whipple (a: ~\cite{whippleA}; b: ~\cite{whippleB}; c: ~\cite{whippleC}),
CANGAROO (a: ~\cite{cangA}; b: ~\cite{cangB}; c: ~\cite{cangC}, d: ~\cite{cangD}),
University of Durham Mark VI Telescope 
(a: ~\cite{durhamA}; b: ~\cite{durhamB}), and Milagro ~\cite{milagro_AGN}.}
         \label{upper_limits}
        \centering
         \begin{tabular}{c c c c c c}
            \hline\hline
            \noalign{\smallskip}
	     Object & $z$ & E$_{\mathrm{th}}$ & I($>$E$_{\mathrm{th}}$) & Crab & Previous Observations\\
             & & [GeV] & [10$^{-12}$ cm$^{-2}$ s$^{-1}$] & \% & \\
            \noalign{\smallskip}
            \hline
            \noalign{\smallskip}
	{\it BL\,Lacs} \\
            \noalign{\smallskip}
	     1ES\,0145+138    & 0.125    & 310 & 2.11 & 1.5 & HEGRA, Whipple$^{\mathrm{c}}$ \\
	     1ES\,0229+200    & 0.140    & 410 & 2.76 & 3.1 & HEGRA, Whipple$^{\mathrm{b,c}}$, Milagro \\
	     1ES\,0323+022    & 0.147    & 210 & 3.92 & 1.5 & HEGRA, Whipple$^{\mathrm{c}}$, Mark VI$^{\mathrm{a}}$, Milagro\\
             PKS\,0548$-$322    & 0.069 & 190 & 6.65 & 2.2 & Whipple$^{\mathrm{a}}$, CANGAROO$^{\mathrm{a,b,d}}$, Mark VI$^{\mathrm{b}}$\\
             EXO\,0556.4$-$3838 & 0.034  & 250 & 10.1 & 5.1 & \\
             RGB\,J0812+026   & ---      & 220 & 7.49 & 3.1 & Milagro\\
             RGB\,J1117+202   & 0.139   & 610 & 1.44 & 3.0 & HEGRA, Whipple$^{\mathrm{b}}$, Milagro\\
	     1ES\,1440+122    & 0.162    & 290 & 5.11 & 3.3 & HEGRA, Milagro\\
             RBS\,1888        & 0.226    & 170 & 3.19 & 0.9 & \\
             Q\,J22548$-$2725   & 0.333  & 170 & 5.83 & 1.6 & \\
             PKS\,2316$-$423    & 0.055  & 190 & 4.13 & 1.4 & Whipple$^{\mathrm{a}}$, CANGAROO$^{\mathrm{a}}$, Mark VI$^{\mathrm{a}}$\\
 	     1ES\,2343$-$151    & 0.226  & 160 & 6.43 & 1.6 & \\
	\\
	{\it Radio-Loud Galaxies} \\
            \noalign{\smallskip}
	     3C\,120          & 0.033   & 230 & 0.92 & 0.4 & HEGRA \\
             Pictor\,A        & 0.034   & 220 & 3.33 & 1.4 & \\
 	     3C\,273          & 0.158    & 280 & 2.90 & 1.8 & HEGRA, Whipple$^{\mathrm{a}}$ \\
             Cen\,A           & 0.00183  & 190 & 5.68 & 1.9 & CANGAROO$^{\mathrm{c}}$, Mark VI$^{\mathrm{a}}$\\
	\\
	{\it Radio-Weak Galaxies} \\
             NGC\,1068        & 0.00379 & 210 & 3.28 & 1.3 & \\
             NGC\,3783        & 0.00965  & 220 & 6.04 & 2.5 & \\
             NGC\,7469        & 0.016    & 250 & 1.27 & 0.6 & \\
            \noalign{\smallskip}
            \hline
       \end{tabular}
   \end{table*}

For the remaining undetected AGN, 99.9\% upper limits on the integral flux 
(assuming a power law spectrum with $\Gamma$=3.0) 
above the energy threshold of the observations, 
and references to previously published limits (when available), 
are shown in Table~\ref{upper_limits}.  The photon index, $\Gamma$=3.0, was chosen
for two reasons:  First, the recently measured VHE spectra of several AGN
(e.g. PKS\,2155$-$304, PKS\,2005$-$489) are softer than the Crab Nebula-like index
of $\Gamma$=2.5 often used for VHE upper limits in past publications.  
Second, the softer index was 
chosen to account for the possible steepening of the observed spectra of the AGN due
to the absorption of $\gamma$-rays on the EBL. Assuming a different photon index 
(i.e. $\Gamma$ between 2.5 and 3.5) has less than a $\sim$10\% effect on the
reported limits, and the systematic error on the upper limits is estimated to
be $\sim$20\%. The percentage of the Crab Nebula flux shown in Table~\ref{upper_limits} is
calculated relative to the integral flux, above the same threshold, determined
from the H.E.S.S. Crab Nebula spectrum.  The H.E.S.S. limits are considerably ($>$5 times) 
stronger than any reported to date.  However, due to the generally variable nature of AGN emission,
these upper limits constrain the maximum average brightness of the AGN only during the observation
time. Hence they are limits on the steady-component or quiescent flux
from the AGN.  Future flaring behavior may increase the VHE flux from any of 
these AGN to significantly higher levels. 

A search for VHE flux variability from each observed AGN was also performed. Here
the nightly integral flux above the average energy threshold was calculated assuming
a photon index of $\Gamma$=3.0 and fit by a constant. 
Any flaring behavior would be demonstrated in the form of a poor $\chi^2$ probability
for the fit.  Table~\ref{VHE_flares} shows the dates each AGN was observed and 
the resulting $\chi^2$ probability. As can be seen, no evidence for VHE flux variability
is found.  

   \begin{table}
      \caption{The dates of the H.E.S.S. observations of each AGN and the $\chi^2$ probability for
a fit of a constant to the nightly integral flux .}
         \label{VHE_flares}
        \centering
         \begin{tabular}{c c c}
            \hline\hline
            \noalign{\smallskip}
	     Object & MJD$-$50000 & P($\chi^2$)\\
            \noalign{\smallskip}
            \hline
            \noalign{\smallskip}
	{\it BL\,Lacs} \\
            \noalign{\smallskip}
	     1ES\,0145+138    & 3202,3205,3210-14 & 0.85\\
	     1ES\,0229+200    & 3317 & - \\
	     1ES\,0323+022    & 2904-05,3267 & 0.87\\
             PKS\,0548$-$322  & 3296,3299-300,3303 & 0.83 \\
             EXO\,0556.4$-$3838 & 3347,3354 & 0.67 \\
             RGB\,J0812+026   & 3081 & - \\
             RGB\,J1117+202   & 3054,3112,3114,3116 & 0.85 \\
	     1ES\,1440+122    & 3110,3119 & 0.13 \\
             RBS\,1888        & 3207-08,3210 & 0.72 \\
             Q\,J22548$-$2725   & 3201,3210-11 & 0.90 \\
             PKS\,2316$-$423    & 3201,3207-08 & 0.50 \\
 	     1ES\,2343$-$151    & 3211-13 & 0.71 \\
	\\
	{\it Radio-Loud Galaxies} \\
            \noalign{\smallskip}
	     3C\,120          & 3316-19,3353-55 & 0.81 \\
             Pictor\,A        & 3269,3318-19,3351,3353-54 & 0.35 \\
 	     3C\,273          & 3109-10,3148-49 & 0.25 \\
             Cen\,A           & 3111-13 & 0.13 \\
	\\
	{\it Radio-Weak Galaxies} \\
             NGC\,1068        & 3290,3292-94,3296 & 0.55 \\
             NGC\,3783        & 3107-08 & 0.52 \\
             NGC\,7469        & 3202,3206,3211-12 & 0.89 \\
            \noalign{\smallskip}
            \hline
       \end{tabular}
   \end{table}

The lack of any significant VHE detection or flaring behavior is
perhaps expected from the beahvior of the individual AGN in the X-ray regime.  
Quick-look results provided by ASM/RXTE team show that none of the AGN (for which
all-sky monitor data exists) were particularly active during the dates of
the H.E.S.S. observations.  On these dates, the measured daily average count rate
from each AGN never deviated by more than $\sim$2$\sigma$ from the mean
value averaged over the whole X-ray data set.

\section{Discussion}

Since AGN are known to emit radiation in all wavebands, understanding and modelling their emission must
take into account their entire spectral energy distribution (SED). Constraining any model 
is difficult as only a limited number of high-energy measurements currently exist
(see ~\cite{EGRET_limits} for EGRET upper limits on blazars, 
Seyfert galaxies and radio-loud galaxies).  This is especially true at VHE energies, 
making the upper limits presented 
in Table~\ref{upper_limits} quite useful due to their unprecedented
strength.  While such modelling is beyond the scope of this paper, 
the applicability and usefulness of the limits for
each of the three classes of observed AGN are discussed.

\subsection{BL\,Lacs}

BL\,Lacs belong to the sub-class of radio-loud AGN known as blazars, which are AGN
thought to possess a jet which is viewed close to the line of sight (\cite{AGN_model}).
The distinction between BL Lacs and other blazars is primarily based
on their optical spectra which are characterized by weak or absent emission lines.
As mentioned in the introduction, almost all VHE bright AGN belong to this class.
These AGN have dominantly non-thermal emission and are characterized by a double-humped SED.  
The low-energy component is widely accepted as originating
from synchrotron radiation of relativistic electrons in the magnetic field around
the object.  However, the origin of the high-energy component is the subject of
much debate.  Various models involving either leptonic or hadronic processes
have been proposed and can be constrained using the H.E.S.S. results.  However,
some caveats are required for interpreting a blazar SED with the H.E.S.S. upper limits.

Blazars are known to be highly variable at all wavelengths, typically characterized
by low-emission quiescent states interrupted by periods of flaring behavior where the
flux increases dramatically.  In some cases this increase is several orders of magnitude.
Due to this extreme variability, it has been shown that fitting the SED of blazars
has very large uncertainties when non-simultaneous multiwavelength (MWL) data are used 
(see e.g.~\cite{SED_fitting}).  As a result the usefulness of non-simultaneous 
upper limits, as is the case for the H.E.S.S. observations, in modelling these object is limited.
The H.E.S.S. upper limits, in the absence of simultaneous MWL data, 
are only relevant for modelling the quiescent state of the blazar using archival 
low-state MWL data.  An additional problem
using these upper limits arises due to the absorption of VHE photons on the EBL.  
The upper limit on the flux intrinsic to the object can be 
significantly higher than those presented in Table~\ref{upper_limits} depending on the redshift.  
As a result the upper limits must have the effects of the EBL removed before they can 
be used for modelling.  Unfortunately, parameterizations of the EBL 
are poorly constrained leading to numerous models with dramatically different behaviors, 
adding another significant uncertainty when using VHE upper limits 
to help model blazar emission.  Given the wide range of EBL interpretations,
this deabsorption is not performed here.

Taking note of the caveats regarding the effects of the EBL and the 
issues with non-simultaneous observations, 
a comparison of the upper limits is made, where possible, to 
three sets of VHE flux predictions based on the SEDs of blazars.  
The first set (\cite{stecker}), referred to as {\it SDS} henceforth, 
uses simple scaling arguments to predict VHE fluxes for Einstein Slew survey objects.
In the case of the {\it SDS} flux predictions the effects of EBL absorption 
are already accounted for with an ''averaged'' model.  The other two sets of predictions
are taken from Costamante \& Ghisellini (2002).  The first, referred to as {\it FOS}, uses a 
phenomenological description of the average SED of blazars based on their bolometric luminosity
~\cite{Fossati}, modified by ~\cite{Donato}, 
and derives predictions on the basis of the individual blazar's 
radio luminosity and synchrotron peak frequency.  The second, referred to as {\it CG},
uses fits of a synchrotron self-Compton
model to existing multiwavelength data.  Both the {\it FOS} and {\it CG}
predictions do not have the effects of EBL absorption accounted for.  This could change the
flux predictions by factors of $\sim$5 above 300 GeV and by factors $>$100 above 1 TeV 
for objects at $z$$\sim$0.2.

Table~\ref{upper_limits2} shows the 99.9\% H.E.S.S. flux upper limits extrapolated 
(assuming $\Gamma=3.0$) to above 300 GeV and above 1 TeV, as well as which predictions are
available above these thresholds.   The H.E.S.S. upper limits are below 
the {\it SDS} predictions above 300 GeV in three of the five cases 
(factors ranging from $\sim$2 to $\sim$5), 
and below two of the five predictions above 1 TeV (factors of $\sim$1.3 and $\sim$5).  
Even if the EBL absorption effects are accounted for in the {\it SDS} predictions, 
the discrepancies can easily be accounted for by the aforementioned simultaneity caveats and thus
the H.E.S.S. upper limits do not make any strong statements regarding the {\it SDS}
predictions.  All the {\it FOS} predictions are above the H.E.S.S. upper limits, 
from factors of $\sim$1.4 to $\sim$16 for the predictions above 300 GeV 
and factors of $\sim$5 to $\sim$40 for the predictions above 1 TeV.  While at first this seems severe,
accounting for TeV absorption can reduce these discrepancies dramatically. In addition
the {\it FOS} predictions are claimed to be more suitable for ''high'' state VHE flux predictions, 
whereas the H.E.S.S. upper limits are most appropriate for constraining the
quiescent state of the AGN.  Given that
variability of up to two orders of magnitude have been seen in VHE blazars such as Markarian\,421,
it is clear that it is again difficult to test these predictions with
the H.E.S.S. upper limits.  However, the disagreement suggests that 
different sets of parameters might be necessary to account for the quiescent state of the source.
The {\it CG} predictions, which are claimed to be more appropriate 
for the quiescent state of the AGN tested by the H.E.S.S. upper limits, 
are all below the upper limits.

   \begin{table}
      \caption{Integral flux upper limits (99.9\% confidence level, $\Gamma$=3.0) 
scaled to above 300 GeV, and above 1 TeV, from H.E.S.S. observations for the
selected AGN where any of the {\it SDS}, {\it FOS} and {\it CG} predictions are available.  
The units are 10$^{-12}$ cm$^{-2}$ s$^{-1}$.  The cases for which the H.E.S.S.
upper limit is below the predicted flux are shown in bold.}
         \label{upper_limits2}
        \centering
         \begin{tabular}{c c c c}
            \hline\hline
            \noalign{\smallskip}
	     Blazar & H.E.S.S. & H.E.S.S. & Pred.\\
		& 300 GeV & 1 TeV & \\
            \noalign{\smallskip}
            \hline
            \noalign{\smallskip}
	     0145+138    & 2.25 & 0.203 & {\it \bf SDS} \\
	     0229+200    & 5.15 & 0.464 & {\it \bf SDS}, {\it \bf FOS}, {\it CG} \\
	     0323+022    & 1.92 & 0.173 & {\it SDS}, {\it \bf FOS}, {\it CG} \\
             0548$-$322  & 2.67 & 0.240 & {\it \bf SDS}, {\it \bf FOS}, {\it CG} \\
             0556$-$384 & 7.03 & 0.632 & {\it \bf FOS} \\
             0812+026   & 4.03 & 0.363 & {\it \bf FOS} \\
             1117+202   & 5.99 & 0.539 & {\it \bf FOS} \\
	     1440+122    & 4.77 & 0.429 & {\it SDS}, {\it \bf FOS}, {\it CG} \\
          \noalign{\smallskip}
            \hline
       \end{tabular}
   \end{table}

\subsection{Other Radio-Loud Galaxies}

Speculation exists for detectable levels of VHE emission from the jets of AGN 
without doppler boosting along the line of sight (see e.g. \cite{Felix_TeV_jets}).
Therefore, the H.E.S.S. observation program also included four other radio-loud AGN.  Like 
BL\,Lacs they all possess jets. One of these, 3C\,273, meets some, 
but not all, of the phenomenological criteria for classification as a blazar.  
However, it is most accurately characterized as a quasar.  It is the brightest and one of the 
most nearby quasars. The other three AGN, Cen\,A, 3C\,120 and Pictor\,A, are found in 
Fanaroff-Riley (FR) radio galaxies. 
These galaxies fall into two classes, FR I and FR II.
The distinction is based on their radio morphology \cite{FR-morph}.
FR I objects, such as Cen\,A (the prototype), 3C\,120 and the VHE-emitter M 87,
show extended jets with no distinct termination point, and
FR II objects, like Pictor\,A, have narrow, collimated jets with terminal ''hotspots.''
These FR objects differ from BL\,Lacs mainly due to a large 
viewing angle (50$^{\circ}-$80$^{\circ}$) with respect to the jet axis.  

Chandra observations (for a review see e.g. ~\cite{Chandra_features}) 
show that Pictor\,A, 3C\,120, and 3C\,273 all possess bright X-ray features like knots and hot spots in 
their large-scale extragalactic jets.  The X-ray fluxes of these features
are at least a factor of 10 larger than the radio and optical fluxes.  
This behavior is the opposite of the predictions from 
synchrotron self-Compton and inverse-Compton models 
and requires alternative theoretical explanations (see e.g. \cite{felix_knots}).
Use of the H.E.S.S. upper limits for these objects should aid in constraining some 
of the presented scenarios.   However, they are still subject to the 
aforementioned variability and EBL absorption (mainly for 3C\,273) caveats.

Located at a distance of 3.4 Mpc, Cen\,A (NGC\,5128) is the closest radio-loud AGN.  
It is one of the best-studied extragalactic objects due to
its large apparent brightness in all wavebands (for a recent review
see ~\cite{CenA}).  The proximity of Cen\,A means that the intrinsic spectrum of the object 
is unaffected by absorption on the EBL, considerably simplifying the use of the H.E.S.S. upper limit
in the modelling of its VHE emission.  However, the lack of simultaneous observations is still an issue as
Cen\,A, like blazars, exhibits large flux variability, albeit on much longer time scales (years). 

During the early days of VHE astronomy, a detection of emission above 300 GeV from Cen\,A was 
claimed using a non-imaging Cherenkov system \cite{Grindlay} during a historically 
high emission state.  The flux reported, I($>$300 GeV) = (4.4$\pm$1.0) $\times$ 10$^{-11}$ 
cm$^{-2}$ s$^{-1}$, is over an order of magnitude above 
the H.E.S.S. 99.9\% flux upper limit extrapolated to above 300 GeV, 
I($>$300 GeV) $<$ 2.3 $\times$ 10$^{-12}$ cm$^{-2}$ s$^{-1}$.
The H.E.S.S. result does not contradict the claimed detection as RXTE ASM observations
show that Cen\,A was in a low emission state when observed by H.E.S.S..  
During a similar low emission state,
EGRET detected $>$100 MeV $\gamma$-ray emission from Cen\,A \cite{EGRET_cenA}. 
This is the only EGRET detection associated with an AGN that is not a member of the blazar class.
Extrapolating the EGRET spectrum to above the H.E.S.S. threshold yields 
I($>$190 GeV) = 3.5 $\times$ 10$^{-12}$  cm$^{-2}$ s$^{-1}$ which is $\sim$60\% of 
the upper limit shown in Table~\ref{upper_limits}.  The H.E.S.S. limit is similar, 
5.5 $\times$ 10$^{-12}$ cm$^{-2}$ s$^{-1}$,  when 
assuming the measured EGRET spectrum of $\Gamma$=2.40.  These results imply that
future identification of a high-emission state in Cen\,A should motivate further VHE observations. 

\subsection{Radio-Weak Galaxies}

All of the radio-weak AGN observed by
H.E.S.S. are located in Seyfert galaxies which differ from the 
galaxies previously discussed in many respects.  Like the other AGN, they
have outflows, albeit typically with low velocity and uncollimated, approximately 
perpendicular to the accretion disk.  Some even have collimated jets that emit synchrotron radiation.  
However, the jets are neither as collimated as in radio-loud
AGN, nor do they show any indications of relativistic motion.   Two kinds of 
Seyfert galaxies (types I and II) exist whose differences can be explained in 
terms of viewing angle \cite{Seyfert}. It is believed that Seyfert I galaxies are viewed ''face on''
and thus the nuclear regions are directly visible, whereas
Seyfert II galaxies are viewed ''edge on'' causing the nuclear regions to be obscured by
material (the torus or warped disk).  Currently, no Seyfert galaxies are known to be VHE emitters.  

Three bright well-studied Seyfert galaxies were observed by H.E.S.S.: 
NGC\,1068, NGC\,3783, and NGC\,7469.  NGC\,1068 (M\,77), the prototypical type II object,
is the brightest and closest known Seyfert galaxy and as such is perhaps the best candidate
for detection of this class at VHE energies.  Here it should be noted that 
since the emission from Seyferts is not beamed, their orientation is not as important as with blazars.
NGC\,3783, a classical type I object, is also one of the brightest and closest 
Seyfert galaxies, and one of the most well studied.  It is also interesting in that exceptionally deep
measurements made using the Chandra X-ray Observatory reveal a fast ($>10^6$  km hr$^{-1}$) 
wind of highly ionized atoms blowing away from the galaxy's suspected 
central black hole ~\cite{chandra_ngc3783}.  NGC\,7469, also type I, is unusual in that it has an 
inner ring of gas very close to the nucleus that is undergoing massive star formation ~\cite{starburst}.

None of these objects were detected and the upper limits shown 
in Table~\ref{upper_limits} are quite constraining.  While Seyfert-type galaxies are not necessarily
expected to emit VHE $\gamma$-rays at observable levels, the H.E.S.S. results easily provide 
constraints for modelling.  This is because these AGN generally show
less variability than blazars, and all the ones observed are 
close enough to only have minimal effects from the absorption of VHE photons on the EBL.  
The H.E.S.S. results could be interpreted as implying that Seyfert-type AGN are 
not significant emitters of VHE photons.  
However, the observed sample and exposure times are small, making it premature 
to rule the class out all together.

\section{Conclusions}

H.E.S.S. observed greater than twenty AGN in 2003 and 2004 as 
part of a campaign to identify new VHE-bright AGN.  Several significant 
detections from this campaign have been reported elsewhere (see Table~\ref{VHE_AGN}
for references).  Results presented here detail the AGN observations for which no significant
excess was found, apart from a marginal signal from the well-known VHE-emitter Markarian\,501.
Despite the limited exposure ($<$8 hours) for each of these AGN, the upper 
limits on the VHE flux determined by H.E.S.S. are the most stringent to date,  
demonstrating the unprecedented sensitivity of the instrument.  Clearly the strength of
the limits makes them quite useful, yet it must again be stressed that any 
interpretation using the H.E.S.S. limits must take into account both the EBL 
and the state of the source using simultaneous data at different wavelengths.

The H.E.S.S. AGN observation program is not complete as many 
proposed candidates have not yet been observed.  Further,
more time is scheduled for observations of some of the AGN presented 
here as part of a monitoring effort for blazars.
H.E.S.S. has already detected $\gamma$-ray emission from four AGN, 
including one (PKS\,2005$-$489) never previously detected in the VHE regime.
Clearly the prospects of finding additional VHE-bright AGN are excellent.

\begin{acknowledgements}
The support of the Namibian authorities and of the University of Namibia
in facilitating the construction and operation of H.E.S.S. is gratefully
acknowledged, as is the support by the German Ministry for Education and
Research (BMBF), the Max-Planck-Society, the French Ministry for Research,
the CNRS-IN2P3 and the Astroparticle Interdisciplinary Programme of the
CNRS, the U.K. Particle Physics and Astronomy Research Council (PPARC),
the IPNP of the Charles University, the South African Department of
Science and Technology and National Research Foundation, and by the
University of Namibia. We appreciate the excellent work of the technical
support staff in Berlin, Durham, Hamburg, Heidelberg, Palaiseau, Paris,
Saclay, and in Namibia in the construction and operation of the
equipment.
\end{acknowledgements}

\end{document}